\documentclass[12pt]{iopart}
\usepackage{dcolumn}% Align table columns on decimal point
\usepackage{bm}% bold math
\usepackage{graphicx}
\usepackage{subfigure}

\begin{document}
\title{Reinforcement-Driven Spread of Innovations and Fads}
\author{P. L. Krapivsky$^1$}
\address{$^1$Department of Physics, Boston University, Boston,
MA,  USA,  02215}
\author{S. Redner$^2$ and D. Volovik$^2$}
\address{$^2$Center for Polymer Studies and Department of
  	Physics, Boston University, Boston, Massachusetts 02215, USA}

\begin{abstract} 
  We investigate how \emph{social reinforcement\/} drives the spread of
  permanent innovations and transient fads.  We account for social
  reinforcement by endowing each individual with $M+1$ possible awareness
  states $0,1,2,\ldots,M$, with state $M$ corresponding to adopting an
  innovation.  An individual with awareness $k<M$ increases to $k+1$ by
  interacting with an adopter.  Starting with a single adopter, the time for
  an initially unaware population that consists of $N$ individuals to adopt
  an innovation grows as $\ln N$ for $M=1$, and as $N^{1-1/M}$ for $M>1$.
  When individuals can abandon the innovation at rate $\lambda$, the
  population fraction that remains clueless about the fad undergoes a phase
  transition at $\lambda_c$; this transition is second order for $M=1$ and
  first order for $M>1$, with macroscopic fluctuations accompanying the
  latter.  The time for the fad to disappear has an intriguing non-monotonic
  dependence on $\lambda$.
\end{abstract}

\maketitle

\section{Introduction}

Disease propagation~\cite{BAM}, the spread of technological
innovations~\cite{CKM,B69,B80,M00,R03}, and outbreaks of social and political
unrest~\cite{G78,L94} are all driven by contagion.  In this work, we
investigate how the mechanism of \emph{social reinforcement\/} affects the
contagion-driven evolution of permanent innovations and transient fads in a
simple agent-based model.  Social reinforcement means that an individual
requires multiple prompts from acquaintances before adopting an innovation.
This mechanism was found to foster the adoption of a desired behavior in a
controlled online social network~\cite{C10}.  Social reinforcement stands in
stark contrast to classical models of epidemics~\cite{BAM}, where a
susceptible individual can become infected by a single exposure to the
infection.  Despite the ubiquity of social reinforcement, this mechanism has
been explored only cursorily in previous studies of contagion
spread~\cite{MK01,DW04,Y09,KKLVB,KN10,OR10,P11}.

In our models, awareness is assumed to have a finite number of possible
states.  We quantify this awareness by a variable that ranges over the $M+1$
values, $0,1,2,\ldots,M$.  We define an individual with awareness 0 as being
susceptible, while an individual moves closer to adopting the innovation as
his/her awareness value increases.  Adoption of the innovation occurs when an
individual reaches the highest awareness value $M$.  The population evolves
by repeated interactions between two random individuals.  In each interaction
with an adopter, someone with awareness $k<M$ advances to awareness $k+1$,
while there are no state changes when two non-adopters interact.  In our
\emph{innovation model}, an innovation is adopted permanently; in our
\emph{fad model}, an adopter abandons the fad at a rate $\lambda$ so that it
eventually becomes pass\'e.  \medskip

\section{Permanent Innovations} 

\begin{figure}[ht]
  \centerline{\includegraphics[width=0.35\textwidth]{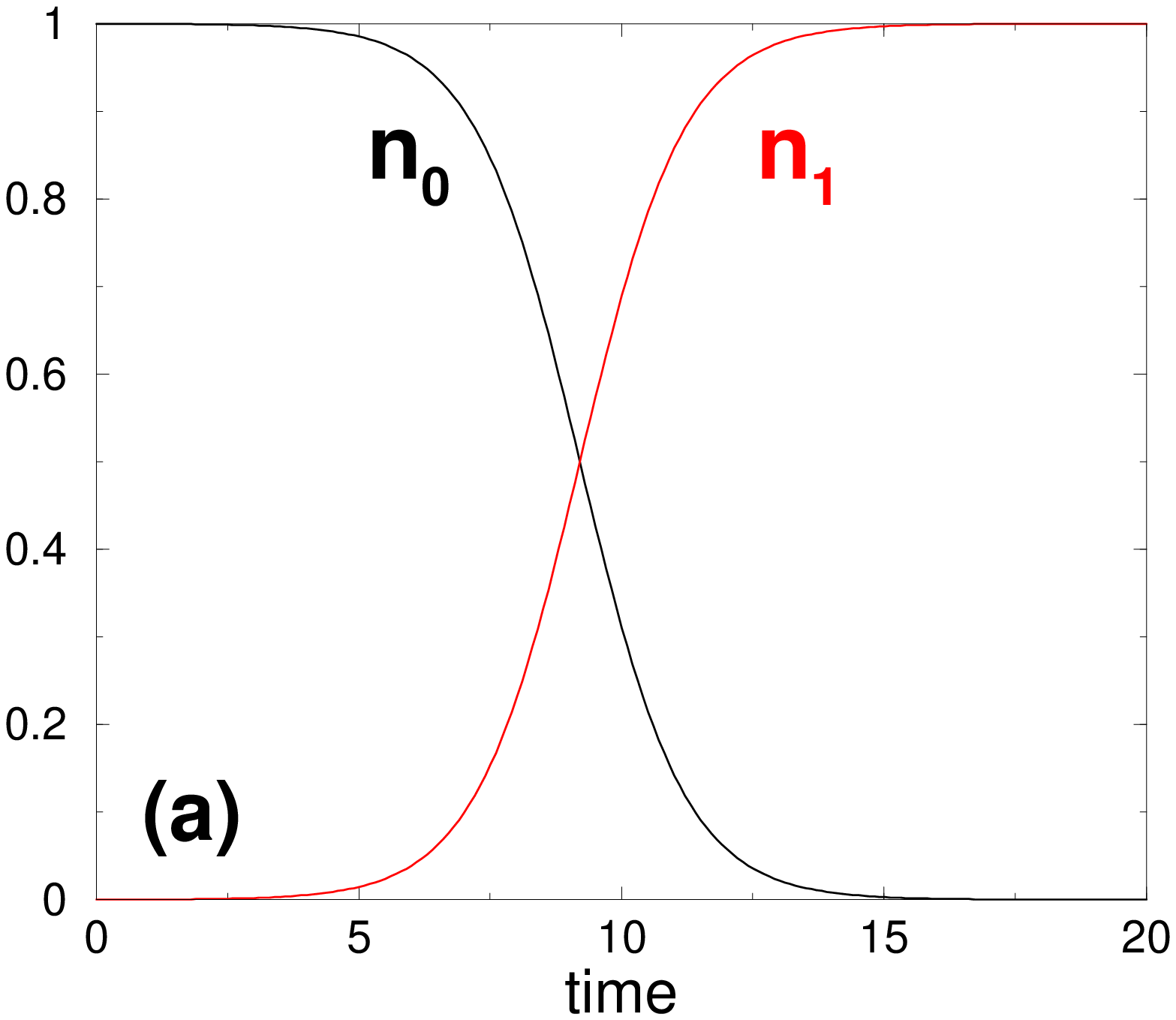}\quad\includegraphics[width=0.35\textwidth]{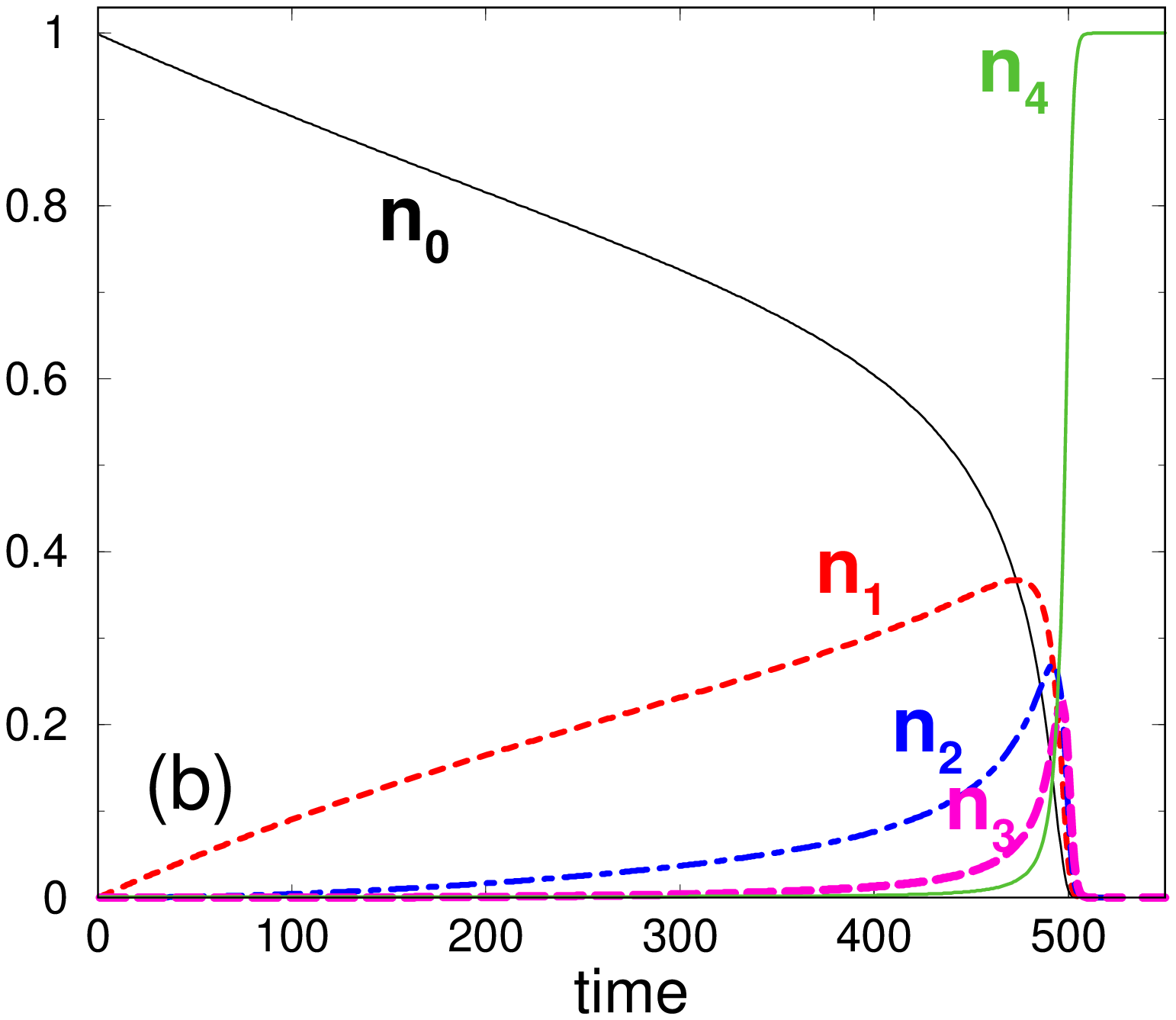}}
\caption{Time dependence of the densities $n_k$ by numerical integration of
  the rate equations for a population of size $N=10^4$, in which
  $n_M(0)=\rho$ and $n_0(0)=1-\rho$, with $\rho=\frac{1}{N}$.  Shown are the
  cases (a) $M=1$ and (b) $M=4$.}
\label{irrev}
\end{figure}

We begin with the simplest situation of no reinforcement
\cite{B69,B80,M00,R03}, namely, a population with two classes of individuals:
susceptible (state 0) and adopters (state 1).  Whenever a susceptible
individual and an adopter meet, the former is converted to an adopter via
$0+1\to 1+1$. The rate equations that give the evolution of a homogeneous and
well-mixed population (the mean-field limit) are:
\begin{equation}
\label{REF}
\dot n_0 = - n_0 n_1, \qquad\qquad \dot n_1 =  n_0 n_1.
\end{equation}
We generically assume that the evolution begins with a small fraction of
adopters in an otherwise susceptible population: $n_1(0)=\rho\ll 1$,
$n_0(0)=1-\rho$.  For this initial condition the solution to the rate
equations is (Fig.~\ref{irrev}) 
\begin{equation}
\label{n1_sol}
n_0=\frac{(1\!-\!\rho)e^{-t}}{\rho+(1\!-\!\rho)e^{-t}}~,
\quad n_1 = \frac{\rho}{\rho+(1\!-\!\rho)e^{-t}}~.
\end{equation}
We define the \emph{emergence time} $t_*$ of the innovation by the criterion
that half of the population has become adopters;
$n_0(t_*)=n_1(t_*)=\frac{1}{2}$ From Eqs.~(\ref{n1_sol}), we have $t_*\simeq
\ln(1/\rho)$.  Ultimately everyone is an adopter; we estimate the resulting
\emph{completion time} $T$ from $n_1(T)=1-\frac{1}{N}$, corresponding to all
but one individual in a population of size $N$ adopting the innovation; this
criterion gives $T \simeq \ln(N/\rho)$~\cite{caveat}.

We now implement social reinforcement by requiring individuals to move to
progressively higher awareness states before adoption ultimately occurs.
Possible examples of such progressions include: not owning a TV, owning a
black \& white TV, owning a color TV~\cite{K11}, or no cell phone, dumb cell
phone, smart phone, etc.  We first treat the simplest example of
reinforcement which is the case of $M=2$.  Here there are three classes of
individuals: susceptible (state 0), aware (state 1), and adopter (state 2),
with respective densities $n_0$, $n_1$, and $n_2$.  In an interaction with an
adopter, a susceptible person becomes aware ($2+0\to 2+1$), while an aware
person adopts the innovation ($2+1\to 2+2$).  All other interactions do not
change individual states. When the rates of all processes are the same, the
rate equations are:
\begin{equation}
\label{12}
\dot n_0 = - n_0 n_2, \quad \dot n_1 =  n_0 n_2 - n_1n_2, \quad \dot n_2=n_1n_2\,.
\end{equation}
To solve these equations we introduce the internal time $\tau=\int_0^t
dt'\,n_2(t')$ to simplify Eqs.~(\ref{12}) to a linear system, whose solution, for
the generic initial condition $n_2(0)=\rho$, $n_1(0)=0$, $n_0(0)=1-\rho$, is
\begin{eqnarray}
\label{n012}
  n_0 &= (1-\rho)\, e^{-\tau}\,,\nonumber\\
  n_1 &= (1-\rho)\, \tau \,e^{-\tau}\,,\\
  n_2 &= 1-(1-\rho)(1+\tau)\, e^{-\tau}\,\nonumber.
\end{eqnarray}

We now define the emergence of the innovation as the point where $n_1$ passes
through a maximum (Fig.~\ref{irrev}(b)). This yields $\tau_*=1$, from which
the corresponding emergence time $t_*$ is given by
\begin{equation}
\label{time:12}
t_* = \int_0^1 \frac{dx}{n_2(x)}=\int_0^1 \frac{dx}{1-(1-\rho)(1+x) e^{-x}}~.
\end{equation}
When $\rho\ll 1$, the asymptotic behavior of the integral is
\begin{equation*}
t_* \simeq \frac{1}{\sqrt{\rho}}\int_0^{1/\sqrt{\rho}} \frac{dy}{1+{y^2}/{2}}
\simeq \frac{\pi}{\sqrt{2\rho}}~,
\end{equation*}
where $y=x/\sqrt{\rho}$, and sub-leading terms are of order of one.  For a
single innovator in a population of size $N$ (corresponding to initial
density $\rho=\frac{1}{N}$), the $N$ dependence of the emergence time is
\begin{equation}
\label{entire}
t_* = \frac{\pi}{\sqrt{2}}\, N^{1/2} + \mathcal{O}(1)\,.
\end{equation}
Thus reinforcement changes the emergence time from a logarithmic to a
power-law $N$ dependence (Fig.~\ref{irrev}).  Using the criterion
$n_2(T)=1-\frac{1}{N}$, we estimate the completion time to be $T =
\pi\sqrt{N/2} + \ln N$ to lowest order~\cite{future}.  Thus once the
innovation emerges, it takes little additional time before it is complete.

For an arbitrary number of intermediate states, an individual with awareness
$k$ increases to $k+1$ by interacting with a adopter, $[M]+[k]\to [M]+[k+1]$,
with $k=0,1,\ldots,M-1$, while all other interactions do not change
individual states.  The corresponding rate equations are
\begin{eqnarray}
\label{gen}
\dot n_0 = - n_M n_0\,,\nonumber\\
\dot n_k = n_M(n_{k-1}-n_k), \quad k=1,\ldots, M-1\,,\\
\dot n_M = n_M n_{M-1}\,\nonumber. 
\end{eqnarray}
By again introducing the internal time $\tau=\int_0^t dt'\,n_M(t')$, we
reduce Eqs.~(\ref{gen}) to a linear system whose solution is
\begin{eqnarray}
n_j = (1-\rho)\frac{\tau^j}{j!}\,e^{-\tau}, \qquad j=0,,\ldots, M\!-\!1\,,\nonumber\\
n_M = 1-(1-\rho)\sum_{j=0}^{M-1}\frac{\tau^j}{j!}\,e^{-\tau}\,.
\end{eqnarray}
In analogy with the case of $M=2$, the innovation emerges at $\tau=1$, where
$n_1$ passes through a maximum (generally, each $n_j$ passes through a
maximum at $\tau=j$).  To obtain explicit time dependences, we must recast
$\tau$ in terms of the physical time via $t = \int_0^\tau dx/n_M(x)$.
Applying the same steps as above and setting $\rho=\frac{1}{N}$, we find the
emergence time
\begin{equation}
\label{entire:M}
t_* = \frac{\pi\,\, (M!)^{1/M}}{M\sin(\pi/M)}\,\, \times N^{1-1/M}~.
\end{equation}
Thus increasing the number of intermediate states $M$ progressively delays
innovation emergence, as the exponent $1-\frac{1}{M}$ approaches 1 as $M$
becomes large (Fig.~\ref{irrev}(c)).  \medskip

\section{Transient Fads}   

\emph{Transient fads} arise when adopters can independently abandon the
innovation at rate $\lambda>0$.  In this case, the innovation can spread to
some degree before it is abandoned and fades away.  The extent of the fad and
its lifetime fundamentally depend on the abandonment rate.  Thus the
population at infinite time consists of adopters who abandoned the fad and
individuals who are forever stuck in intermediate awareness states because of
the absence of catalyzing adopters.  Of particular interest are the
\emph{clueless} individuals who were never exposed to the fad while it was
active.  Their fraction, defined as $c_\infty(\lambda)\!\equiv\!
n_0(t\!\!=\!\!\infty)$, characterizes the competing influences of contagion
and fad abandonment~\cite{rumor}.  For an infinite population, $c_\infty$
undergoes a continuous transition as a function of $\lambda$ for $M=1$, but a
\emph{discontinuous} transition for $M\geq 2$.  Moreover, the time to reach
the final state varies non-monotonically with $\lambda$.

\begin{figure}[ht]
\centerline{\includegraphics[width=0.35\textwidth]{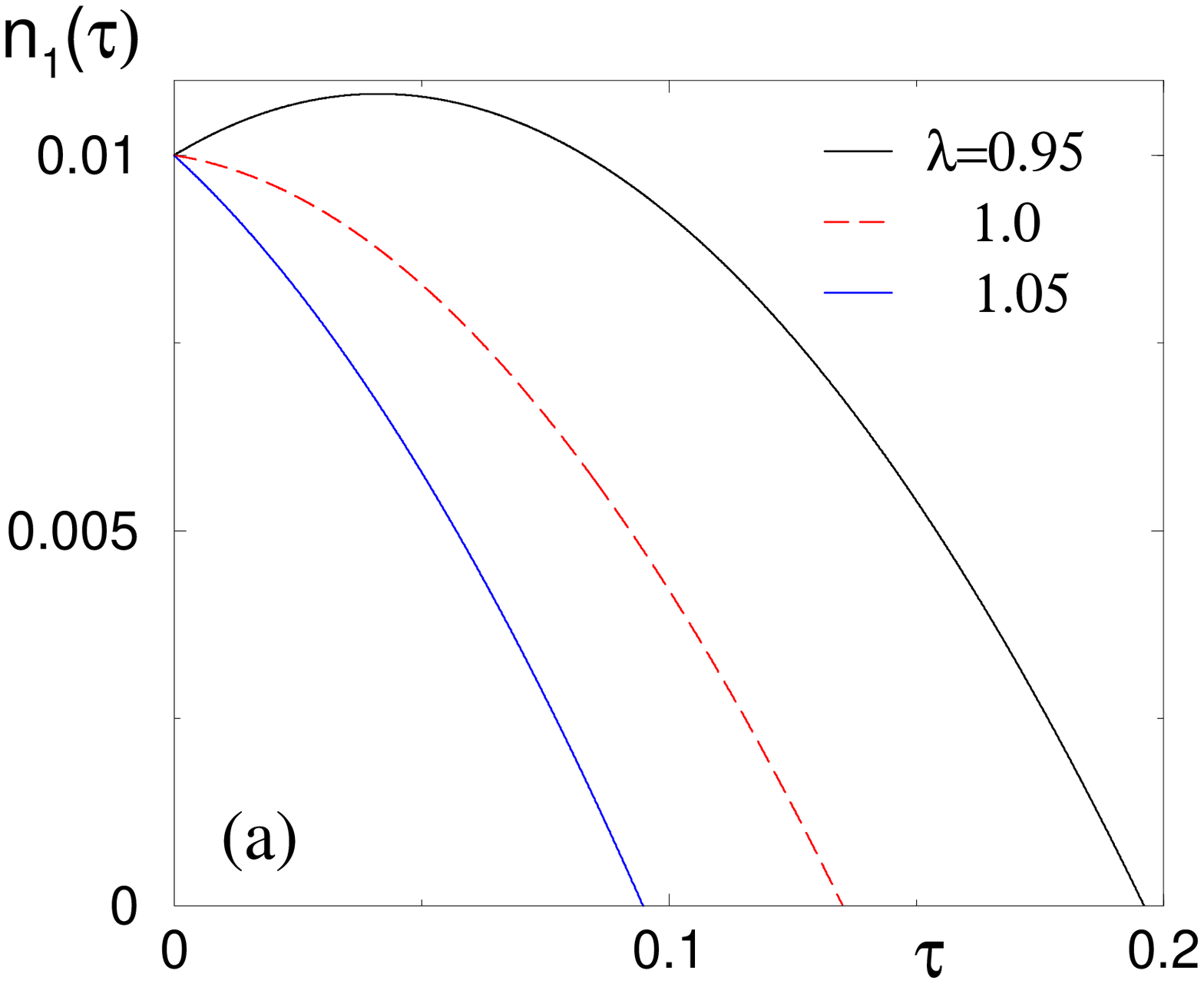}\qquad
\includegraphics[width=0.35\textwidth]{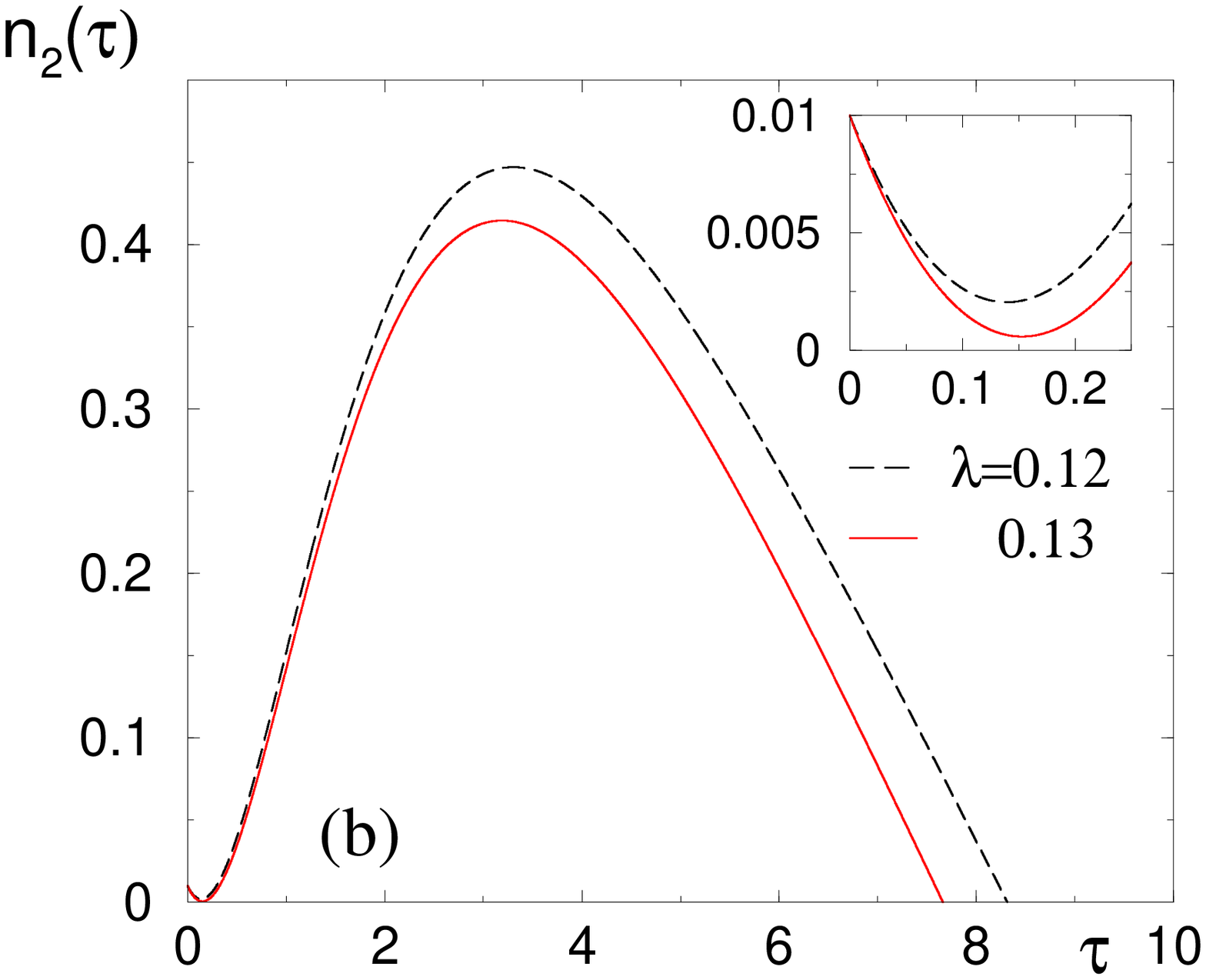}}
\caption{Dependence of $n_M(\tau)$ versus $\tau$ for: (a) no reinforcement
  ($M=1$), and (b) reinforcement, with one intermediate state ($M=2$),
  Eqs.~(\ref{n01_sol}) and (\ref{n2decay}) respectively.  The inset in (b)
  shows the near tangency of $n_2(\tau)$ versus $\tau$ for $\lambda=0.13$.}
\label{n2-tau}
\end{figure}

\begin{figure}[ht]
\centerline{\includegraphics[width=0.5\textwidth]{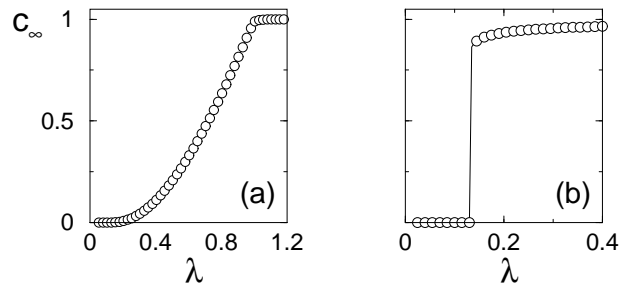}}
\caption{Clueless fraction $c_\infty$ versus abandonment rate $\lambda$ for:
  (a) two-state and (b) three-state models.  The initial adopter fraction is
  $n_1(0)=10^{-4}$ in (a) and $n_2(0)=10^{-2}$ in (b).}
\label{cinf}
\end{figure}

The case of no reinforcement coincides with the classic SIR epidemic
model~\cite{BAM} with the identifications: adopter $\leftrightarrow$
infected, abandoner $\leftrightarrow$ recovered, while the meaning of
susceptible is the same in both models.  The rate equations are $\dot n_0 = -
n_0 n_1$, $\dot n_1 = n_0 n_1-\lambda n_1$, with solution
\begin{equation}
\label{n01_sol}
n_0 = (1-\rho)\, e^{-\tau}\,, \quad n_1 = 1-\lambda\tau -(1-\rho)\, e^{-\tau}~,
\end{equation}
where $\tau = \int_0^t dt'\, n_1(t')$.  The evolution ceases at an internal
stopping time $\tau_\infty$ defined by $n_1(\tau_\infty)=0$; this corresponds
to physical time $t=\infty$.  The condition $n_1(\tau_\infty)=0$ leads to
three regimes of behavior for the clueless fraction $c_\infty$
(Fig.~\ref{n2-tau}(a)).  For $\lambda<1$ (subcritical), adopters abandon the
fad sufficiently slowly that the fad can spread globally before dying out.
In the supercritical regime of $\lambda>1$, adopters abandon the fad so
quickly that there little time for the innovation to spread before it is
extinguished.  In this limit, Eq.~(\ref{n01_sol}) gives
$\tau_\infty=\rho/(\lambda - 1)$ and $c_\infty = 1 - {\rho}/({\lambda -1})$
to leading order, while for $\lambda=\lambda_c=1$, $c_\infty = 1 -
\sqrt{{2\rho}}$.  Thus $c_\infty$ undergoes a continuous transition (in the
$\rho N\gg 1$ limit) as $\lambda$ passes through the critical value
$\lambda_c=1$ (Fig.~\ref{cinf}(a)).

Let us now investigate the role of reinforcement on the dynamics of a fad.
We consider the simplest situation of a single intermediate state; that is,
$M=2$, or equivalently, three internal states for each individual.  In this
case, the evolution of $n_0$ and $n_1$ are again given by Eq.~(\ref{n012}),
while the solution of $n_2$ is
\begin{equation}
\label{n2decay}
n_2 = 1-(1-\rho)(1+\tau) e^{-\tau} - \lambda \tau\,.
\end{equation}
A curious feature of this result is that the density of fad adopters
$n_2(\tau)$ can first decrease, then increase later on, before ultimately
vanishing (Fig.~\ref{n2-tau}(b)).  This unusual behavior stems from the
delicate interplay between abandonment of the fad and the creation of new
adopters from the remaining reservoir of susceptible individuals.  As a
result of the two extrema in $n_2$ as a function of $\tau$, the stopping
condition $n_2(\tau_\infty)=0$ can have one, two, or three roots, depending
on $\lambda$.  This change in the number of roots ultimately causes the
discontinuity in the clueless fraction $c_\infty$ as a function of $\lambda$.

To locate this transition in the supercritical case equation, notice that
(\ref{n2decay}) has three roots.  We are interested in the smallest root and
therefore expand the left-hand side of Eq.~(\ref{n2decay}) for small $\tau$.
Keeping the leading terms gives
\begin{equation}
\label{quadratic}
n_2(\tau_\infty)\approx \rho+\frac{1}{2}\tau_\infty^2 - \lambda \tau_\infty=0\,.
\end{equation}
From this quadratic equation, we see that the interesting behavior arises
when $\lambda = \mu\sqrt{\rho}$ where $\mu=\mathcal{O}(1)$.  With this
convenient parameterization, the solution for $\tau_\infty$ is $\tau_\infty
=\sqrt{\rho}\, [\mu \pm \sqrt{\mu^2-2}]$.  Using the physically relevant
smaller solution, we find, for $\mu>\mu_c$ (which equals $\sqrt{2}$ to lowest
order)
\begin{equation}
 c_\infty = (1-\rho)\,e^{-\tau_\infty} \simeq 1 - \sqrt{\rho} \big(\mu-\sqrt{\mu^2-2}\big);
\end{equation}
i.e., the clueless fraction is close to one (Fig.~\ref{cinf}(b)).  In the
subcritical case, $\mu<\sqrt{2}$, the relevant root of $n_2(\tau_\infty)=0$
is $\tau_\infty = 1/(\mu\sqrt{\rho})$ to leading order.  The clueless
fraction is
\begin{equation}
\label{n0_inf}
c_\infty =  e^{-\tau_\infty}= e^{-1/(\mu\sqrt{\rho})}~,
\end{equation}
which is close to zero for $\rho\to 0$.  Thus the clueless fraction undergoes
a first-order transition as a function of $\lambda$.

\section{Fad Completion Time}

\begin{figure}[ht]
\centerline{\includegraphics[height=0.325\textwidth,width=0.375\textwidth]{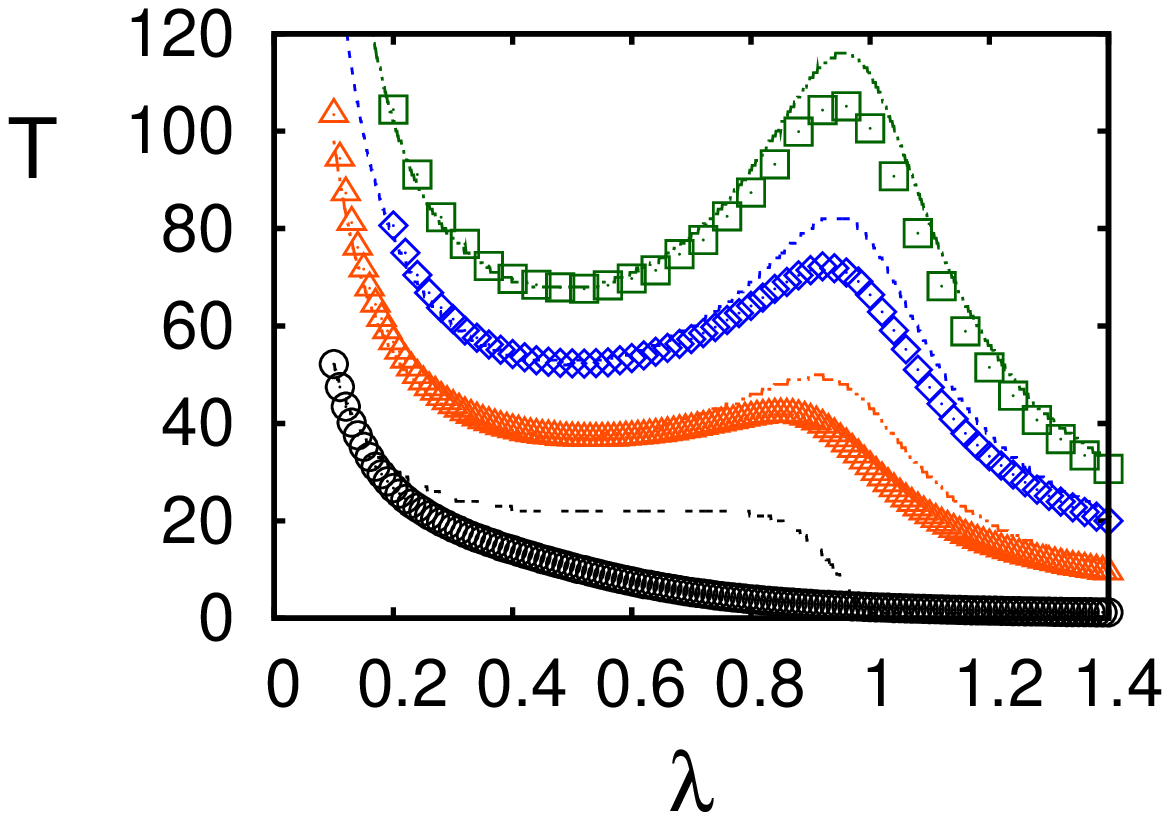}
\includegraphics[height=0.325\textwidth,width=0.375\textwidth]{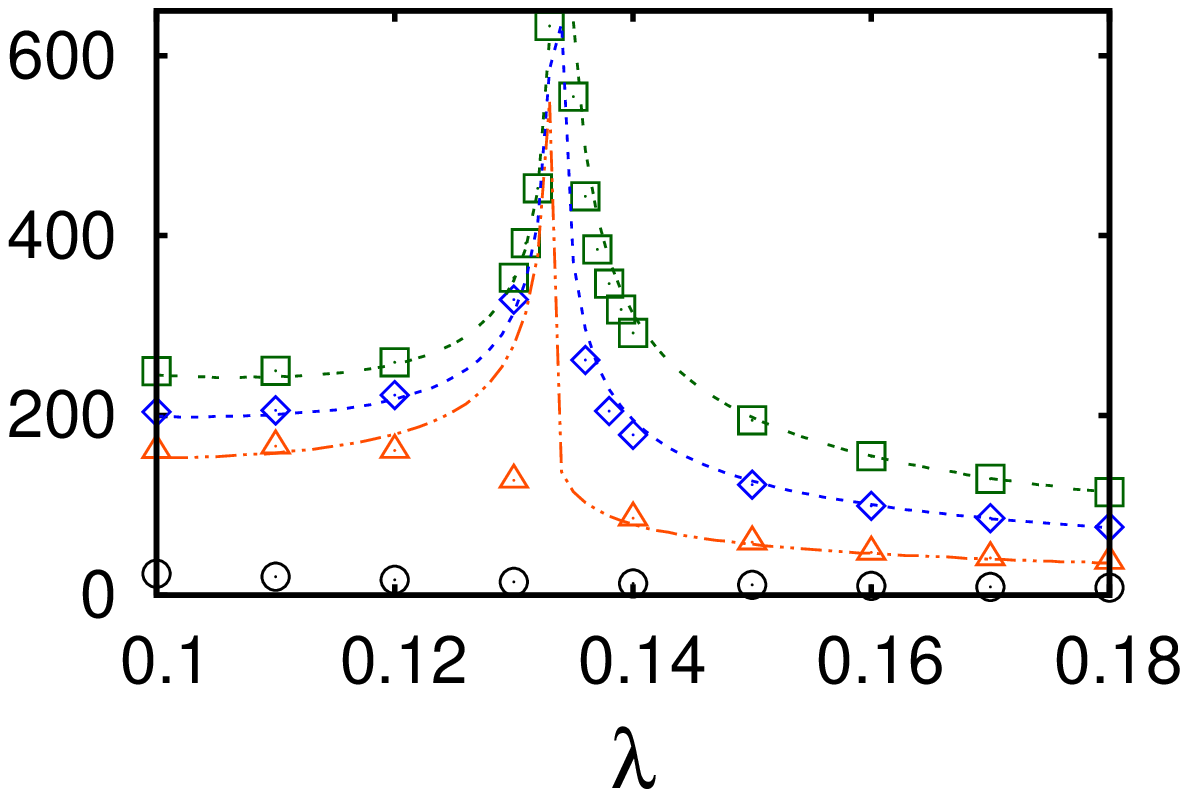}}
\caption{Completion time $T$ versus $\lambda$ for the fad model, for $M=1$
  (left) and $M=2$ (right) with $\rho=10^{-2}$.  Points are simulation
  results for $N=10^2, 10^4, 10^6, 10^8$ (bottom to top).  Dashed lines are
  the corresponding results from numerically integrating the rate equations.
  For $M=1$, the width of the peak at $\lambda_c=1$ scales as $\sqrt{\rho}$.}
\label{tstar}
\end{figure}

A striking aspect of our fad model is that the time for a fad to die out has
a non-monotonic dependence on the abandonment rate $\lambda$
(Fig.~\ref{tstar}).  This non-monotonicity has a simple qualitative origin.
If the abandonment rate is large, then the initial adopters abandon before
they can recruit new adopters.  Thus the fad quickly disappears.  Conversely,
if the abandonment rate is small, essentially the entire population adopts
the innovation \emph{en masse}, after which the fad disappears in a time that
scales as ${1}/{\lambda}$.  Between these two limits, the fad ``smolders''
rather than just extinguishing itself immediately or suddenly igniting and
then quickly extinguishing itself.  In this intermediate range of $\lambda$
values, new adopters are slowly replenished at nearly the same rate as other
adopters abandon the fad, so that the fad can be extremely long lived.

In a population of size $N$, we determine the time for a fad to end not by
the criterion $\tau=\tau_\infty$, where the number of adopters vanishes, but
rather by $n_M(\tau^*)=\frac{1}{N}$.  Namely, only a single adopter remains
in a finite population.  This internal time corresponds to the value $T =
\int_0^{\tau^*} {d\tau}/{n_M(\tau)}$ for the physical time at which the fad
disappears. The actual determination of the completion time is very different
for the cases $M=1$ and $M>1$, and we investigate these two cases in turn.

\subsection{No reinforcement, $M=1$}

Generically, $T$ is proportional to $\ln N$ because $n_1$ goes to zero with a
finite slope as $\tau$ approaches $\tau_\infty$ (Fig.~\ref{n2-tau}(a)).  As a
consequence, the integral for $T$ is logarithmically divergent in $N$.
However, the details of this dependence depends on the value of the
abandonment rate $\lambda$.

In the subcritical regime ($\lambda<1$), we determine $T$ by expanding $n_1$
about $\tau_\infty$ and using the condition $e^{-\tau_\infty} +
\lambda\tau_\infty = 1$ to obtain
\begin{equation}
\label{int_11}
T = \frac{1}{\lambda+\lambda\tau_\infty-1}\int_{1/N} \frac{d y}{y}
 = \frac{\ln N}{\lambda+\lambda\tau_\infty-1}~.
\end{equation}
The lower limit in Eq.~(\ref{int_11}) follows from the stopping criterion
$n_1(\tau^*) = \frac{1}{N}$, while the upper limit is immaterial for the
asymptotic behavior.

In the supercritical regime ($\lambda>1$), the density of adopters $n_1$
decreases almost linearly in $\tau$ over the entire range for which $n_1$ is
positive.  In this case~\cite{rN}, an expansion of $n_1$ about $\tau_\infty$
leads to $T = {\ln(\rho N)}/({\lambda-1})$.  

In the critical case of $\lambda=1$, $n_1$ decreases quadratically with
$\tau$ and the same expansion procedure as outlined above gives $T =
{\ln(\rho N)}/{\sqrt{2\rho}}$ in the asymptotic limit.  Consistent with the
logarithmic dependence at the critical point for the case $\rho=1/N$, we find
completion time distribution has a power law tail, $P(T)\sim T^{-2}$.  The
resulting average completion time is $T=\frac{1}{3}\log{N}$, a behavior that
was obtained previously in the context of epidemic dynamics~\cite{BK04}.

\subsection{Reinforcement, $M>1$}

In this case, the fad evolution in the supercritical regime closely mirrors
the behavior of the $M=1$ case.  In particular, the time dependence of
$n_M(\tau)$ is similar to $n_1(\tau)$ in the case of no reinforcement: $n_M$
approaches zero with finite slope, from which the ending time of the fad
again scales as $\ln N$.  However, in constant to the case $M=1$, the value
of $\tau$ where $n_M(\tau)$ first reaches zero changes discontinuously as
$\lambda$ passes through $\lambda_c$ (Fig.~\ref{n2-tau}(b)).  More
interestingly, when $\lambda\approx \lambda_c$, $n_M$ approaches zero with a
quadratic minimum, as illustrated in the inset to Fig.~\ref{n2-tau}(b).  This
property leads to an algebraic, rather than a logarithmic, dependence of the
completion time on $N$.  Finally for $\lambda<\lambda_c$, $n_M$ again reaches
zero with a finite slope, leading to a logarithmic dependence of the ending
time on $N$.  Thus the time for the fad to disappear at the critical point is
much larger than the corresponding ending times away from this point.  Monte
Carlo simulations of the fad dynamics in a finite population mirror our
analytic predictions, except near the first-order transition, where large
fluctuations arise.

Let us now focus on the properties of the completion time at the first-order
transition point where fluctuations are particularly strong.  There are two
independent and natural scenarios for which to define the lifetime of the
fad: (i) a fixed \emph{number} of initial adopters (generally we treat the
case of one adopter) or (ii) a fixed \emph{fraction} $\rho$ of initial
adopters.  To find the fad lifetime in the former case of $\rho=\frac{1}{N}$,
it is again convenient to use parameterization $\lambda=\mu\sqrt{\rho}$
because the critical value of $\mu$ is $\mathcal{O}(1)$.  We therefore
substitute the critical value $\mu_c=\sqrt{2}$ (to lowest order) into the
expansion (\ref{quadratic}) for $n_2$ to obtain $n_2=\frac{1}{2}(\sqrt{2\rho}
- \tau)^2$.  The ending time for the fad is now given by
\begin{equation}
\label{T}
T = 2\int_0^{\tau^*} \frac{d\tau}{(\sqrt{2\rho} - \tau)^2}~,
\end{equation}
with $\tau*$ determined from the criterion $n_2(\tau^*)=\frac{1}{N}$.  The
latter gives $\tau^*=\sqrt{2\rho}-\sqrt{2/N}$ and using this upper limit in
(\ref{T}) gives $T=\sqrt{2N}$.  However, the prefactor arises from the
imprecise criterion $n_2(\tau^*)=\frac{1}{N}$, and simulations instead give
$T\sim 0.56\,\sqrt{N}$.  For a fixed fraction of initial adopters $\rho$, our
simulations show that the average fad lifetime grows with $N$ roughly as
$N^{1/4}$ for $M=2$, a result for which we do not yet have an explanation.

\begin{figure}[ht]
\centerline{\includegraphics[width=0.55\textwidth]{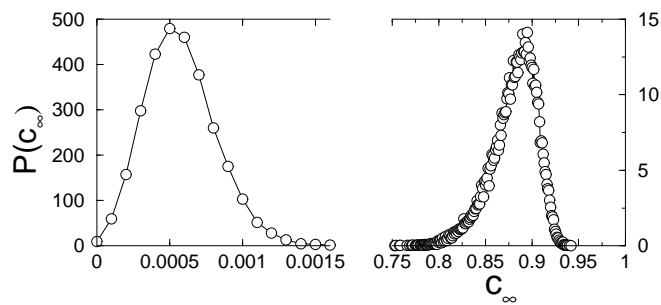}}
\caption{Probability density $P(c_\infty)$ for the fraction of clueless
  individuals at the end of the process at the critical point,
  $\lambda_c=\sqrt{2\rho}$, for the $M=2$ fad model, with initial density
  $\rho=10^{-2}$ and population size $N=10^4$.  (Note the horizontal scale
  break.)}
\label{bimodal}
\end{figure}

\begin{figure}[ht]
\centerline{\includegraphics[height=0.325\textwidth,width=0.375\textwidth]{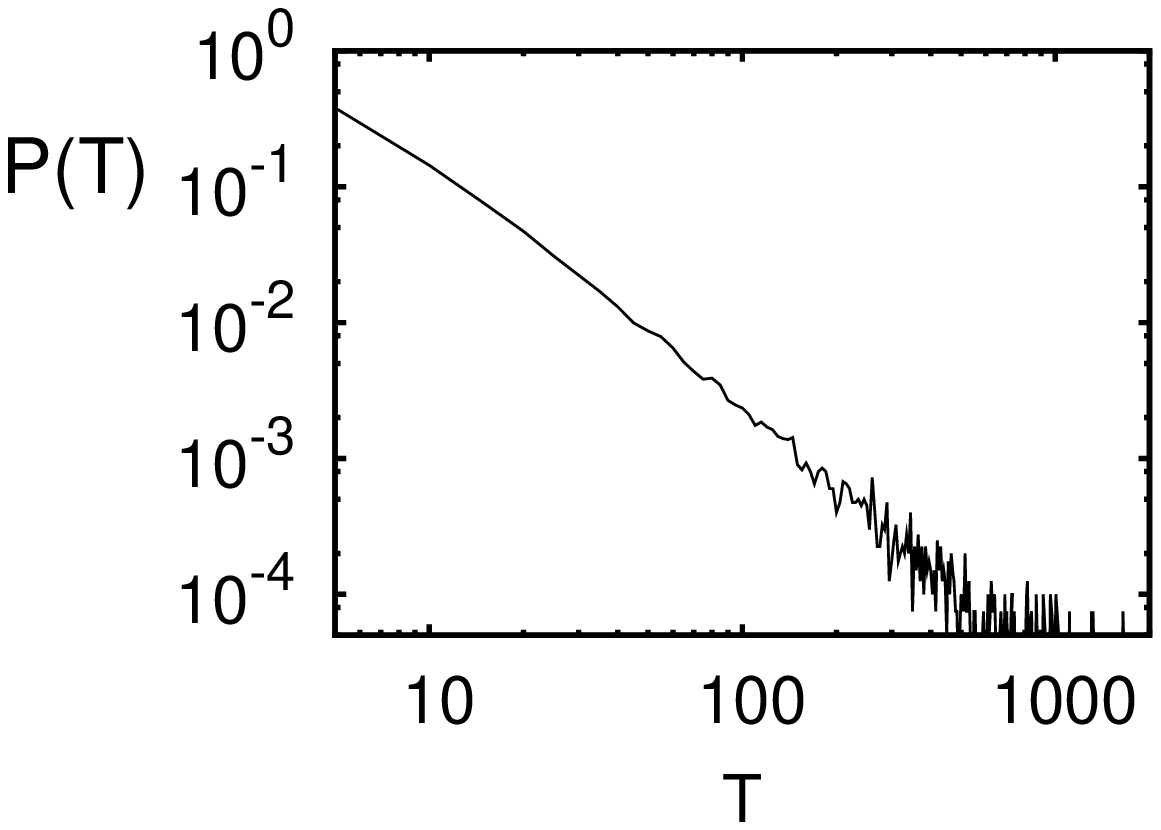}
\includegraphics[height=0.325\textwidth,width=0.375\textwidth]{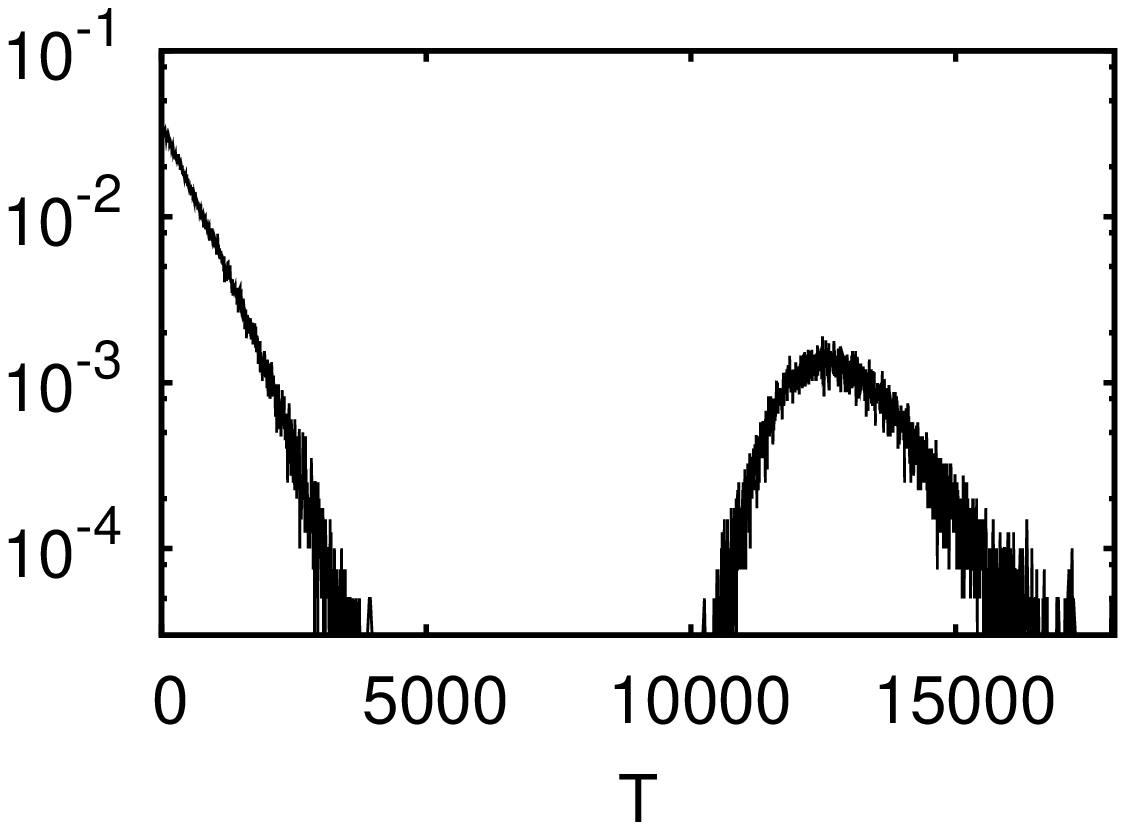}}
\caption{Probability distribution of completion time $T$ for the fad model at
  the critical point, for $M=1$ (left) and $M=2$ (right) with $\rho=1/N$.  We
  use $N=10^9$ for $M=1$ and $N=10^6$ for $M=2$.}
\label{tHist}
\end{figure}

As a result of the large fluctuations near the transition, the completion
time distribution consists of distinct components.  One component corresponds
to realizations where the fad quickly dies out so that the population is
almost entirely clueless (Fig.~\ref{bimodal}).  In contrast, for the
remaining fraction of realizations, nearly everyone adopts and then abandons
the fad.  Corresponding to this dichotomy in the fate of individual
realizations, the distribution of times at which the fad disappears has
distinct short-lived and long-lived contributions (Fig.~\ref{tHist}).

\section{Summary}

We have shown how the mechanism of social reinforcement plays a strong role
in determining how permanent innovations and transient fads are adopted in a
socially-interacting population.  For permanent innovations, we modeled the
effect of reinforcement by endowing each individual with $M+1$ levels of
awareness $0,1,2,\ldots, M$.  An individual increases his/her level of
awareness by one unit as a result of interacting with an adopter, and
adoption occurs when an individual reaches the highest awareness level $M$.
In the mean-field limit, we found that the time for the innovation to be
adopted universally scales as $N^{1-1/M}$, so that increasing $M$ delays the
onset of the innovation.

For transient fads, the fad quickly disappears for $\lambda>\lambda_c$, while
for $\lambda<\lambda_c$ the fad is nearly universally adopted before finally
being forgotten.  The fad lasts the longest when $\lambda=\lambda_c$.  Here
new adopters are slowly replenished as others abandon, so that the fad slowly
smolders rather than igniting and quickly burning out.  The transition in the
fraction of clueless individuals --- those who have no knowledge of the fad
before it disappears --- is second order when there is no reinforcement, but
first order with reinforcement.  The rich phenomenology near the transition
may offer opportunities to help predict the reach of a technological
innovation before it is released on the market.  A fruitful direction for
additional research is to include the effect of stochastic fluctuations and
heterogeneous social connections.  Both of these attributes can be
anticipated to considerably enrich the dynamics that have been uncovered in
this work.

\ack We thank Damon Centola for helpful discussions, Sam Bowles for
literature advice, and NSF grants CCF-0829541 (PLK) and DMR-0906504 (DV and
SR) for financial support.


\begin{thebibliography}{99}
\bibitem{caveat} If one uses the criterion that the innovation is complete
  when $n_1=1$, one obtains the unphysical result that the completion time is
  infinite.
\bibitem{rN} Here we assume $\rho\ll 1$ and $\rho N\gg 1$.  In the extreme
  case of $\rho=N^{-1}$, fluctuations are artificially large and the rate
  equation approach becomes invalid.
\end{thebibliography}
\end{document}